\documentclass[preprint2]{aastex}

\shorttitle{Pixel Analysis of NGC 5195}

\shortauthors{Lee et al.}
\def\simlt{\lower.5ex\hbox{$\; \buildrel < \over \sim \;$}}
\def\simgt{\lower.5ex\hbox{$\; \buildrel > \over \sim \;$}}

\begin{document}

\title{\emph{HUBBLE SPACE TELESCOPE} PIXEL ANALYSIS OF THE INTERACTING S0 GALAXY NGC 5195 (M51B)}

\author{Joon Hyeop Lee, Sang Chul Kim, Chang Hee Ree, Minjin Kim, Hyunjin Jeong, Jong Chul Lee, Jaemann Kyeong}
\affil{Korea Astronomy and Space Science Institute, Daejeon 305-348, Korea}

\email{jhl@kasi.re.kr, sckim@kasi.re.kr, chr@kasi.re.kr, mkim@kasi.re.kr, hyunjin@kasi.re.kr, jclee@kasi.re.kr, jman@kasi.re.kr}

\begin{abstract}
We report the properties of the interacting S0 galaxy NGC 5195 (M51B), revealed in a pixel analysis using the {\it HST}/ACS images in the F435W, F555W and F814W ($BVI$) bands. We analyze the pixel color-magnitude diagram (pCMD) of NGC 5195, focusing on the properties of its red and blue pixel sequences and the difference from the pCMD of NGC 5194 (M51A; the spiral galaxy interacting with NGC 5195). The red pixel sequence of NGC 5195 is redder than that of NGC 5194, which corresponds to the difference in the dust optical depth of $2<\Delta\tau_V<4$ at fixed age and metallicity. The blue pixel sequence of NGC 5195 is very weak and spatially corresponds to the tidal bridge between the two interacting galaxies. This implies that the blue pixel sequence is not an ordinary feature in the pCMD of an early-type galaxy, but that it is a transient feature of star formation caused by the galaxy-galaxy interaction. We also find a difference in the shapes of the red pixel sequences on the pixel color-color diagrams (pCCDs) of NGC 5194 and NGC 5195. We investigate the spatial distributions of the pCCD-based pixel stellar populations. The young population fraction in the tidal bridge area is larger than that in other areas by a factor $> 15$. Along the tidal bridge, young populations seem to be clumped particularly at the middle point of the bridge. On the other hand, the dusty population shows a relatively wide distribution between the tidal bridge and the NGC 5195 center.
\end{abstract}


\section{INTRODUCTION}

NGC 5195 is an SB0 galaxy \citep{san87} interacting with the face-on spiral galaxy NGC 5194. Since this interacting system (NGC 5194 + NGC 5195; also called as M51) is very nearby from us and has a face-on inclination, the two interacting galaxies are very good targets to study the effects of tidal interaction on the stellar populations and interstellar medium of the galaxies.
In this reason, since Lord Rosse presented the nice sketch of M51 in the middle of 19th century, many studies have been carried out about M51 in various wavelengths and methods.
The galaxy interaction must largely affect the distribution of interstellar medium, which can be studied through infrared and radio observations.
\citet{smi82} estimated the far-infrared luminosity of this system and the contribution of NGC 5195 (estimated to be $10^{10} L_{\odot}$, about one third of the total far-infrared luminosity).
Later, \citet{rot90} investigated the H {\scriptsize I} morphology of M51, such as warp, oval distortion, streaming motion along the spiral arms, tidal tails and bridges.
More recently, CO imaging studies revealed that the total molecular gas masses of NGC 5194 and NGC 5195 are $4.9\times10^9 M_{\odot}$ and $7.8\times10^7 M_{\odot}$ respectively \citep{kod11}, and that the molecular gas in the central region of NGC 5195 is gravitationally too stable to form the dense cores \citep[which is the reason why the massive star formation in NGC 5195 seems to be suppressed in spite of the abundant molecular gas;][]{koh02}.

While the infrared and radio studies are useful to study the properties of interstellar medium, studies in the optical bands are necessary to understand the properties of stellar populations.
\citet{dob10} reproduced the present day spiral structure of NGC 5194 and many details of the tidal interaction between NGC 5194 and NGC 5195 in hydrodynamic simulations, finding that at most times there is no offset between the stars and gas within errorbars. That is, it is inferred that the tidal interaction must have affected the stellar populations as well as the interstellar medium in M51.

The studies of M51 in the optical bands also have a very long history. For example, \citet{oka76} and \citet{bur78} conducted surface photometry of NGC 5194 and NGC 5195, revealing their basic photometric structures. However, the studies of stellar light in M51 were relatively less active than those of gas or dust in M51.
An important chance activating the optical studies on M51 was the observation of it using the {\it Hubble Space Telescope} ({\it HST}).
Recently, the very large area of the interacting system M51 was covered using the {\it HST} and Advanced Camera for Surveys (ACS) with very high resolution by the Hubble Heritage Team \citep{mut05}.
This high-quality data (combined with several other data sets) enabled many interesting studies on M51: the ages and spatial distribution of their star clusters \citep{lee05,hwa08,hwa10}, the properties of bright stars \citep{kal10}, the luminosity function of H{\scriptsize II} regions \citep{lee11a}, the effects of interaction \citep{dur03,mei08}, and so on. Those previous studies analyzed the entire system of M51 or mainly focused on the properties of NGC 5194, but studies focusing on the properties of NGC 5195 are relatively rare.

One of such rare studies is the study of faint fuzzy clusters in NGC 5195 by \citet{hwa06}. They found about 50 faint fuzzy star clusters around NGC 5195 that are redder and larger than typical globular clusters. Interestingly, most of those clusters are scattered in an elongated region almost perpendicular to the northern spiral arm of NGC 5194, and the center of the region is slightly north of the NGC 5195 center, while normal compact red clusters of NGC 5195 are located around the bright optical body of the host galaxy. \citet{hwa06} suggested that at least some faint fuzzy clusters are experiencing tidal interactions with the companion galaxy NGC 5194 and must be associated with the tidal debris in the western halo of NGC 5195, which means that the tidal interaction between galaxies directly affects the stellar properties. However, using stellar photometry, only star clusters or very bright stars can be studied, because most stars in NGC 5195 are not resolved in our current image data.

One powerful approach to the stellar population analysis for the objects that are not close enough to resolve their individual stars (such as NGC 5195) is the {\it pixel analysis} \citep[e.g.][]{kas03,deg03,lan07}. In this method, the individual pixels (instead of stars) are used as a unit for the stellar population analysis. The pixel color-magnitude diagrams (pCMDs) and pixel color-color diagrams (pCCDs) provide constraints on the stellar properties like age, metallicity and dust distributions and some collective properties of target galaxies such as galaxy morphology \citep[][hereafter Paper~I]{lee11b}.

In Paper~I, we used the pixel analysis method to investigate the stellar properties of NGC 5194, finding that the pCMD of NGC 5194 shows two major pixel sequences: red pixel sequence and blue pixel sequence, which represent old ($>1$ Gyr) stellar population in the bulge and young ($<1$ Gyr) stellar population in the disk, respectively.
The color variation of the red pixel sequence corresponds to a metallicity variation of $\Delta$[Fe/H] $\sim2$ or an optical depth variation of $\Delta\tau_V\sim4$ by dust, but the actual color variation is thought to be the combination of the two effects. The color variation of the blue pixel sequence corresponds to age variation from 5 Myr to 300 Myr.
We also found that the pixel population distributions agree with a compressing process by spiral density waves. In addition, the tidal interaction between NGC 5194 and NGC 5195 appears to enhance the star formation at the tidal bridge connecting the two galaxies.

This paper presents a pixel analysis of the interacting S0 galaxy NGC 5195, subsequent to Paper~I.
The main goal of this paper is to improve the understanding of NGC 5195 properties using the pixel analysis method, particularly focusing on the effects of tidal interaction on the stellar properties.
The outline of this paper is as follows. Section 2 describes the data and our pixel analysis. The results of the pixel analysis are shown in Section 3 and discussed in Section 4. Section 5 gives the conclusion. Throughout this paper, we adopt
the cosmological parameters: $h=0.7$, $\Omega_{\Lambda}=0.7$, and
$\Omega_{M}=0.3$.

\section{DATA AND ANALYSIS}

\begin{figure}[!t]
\plotone{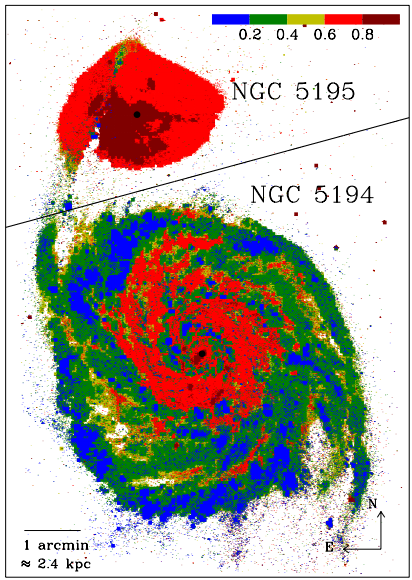}
\caption{ $B-V$ colormap of M51 (NGC 5194 + NGC 5195), using pixels brighter than $V<21.5$ mag arcsec$^{-2}$. The solid line is the 7:3 division between the NGC 5194 and NGC 5195 centers and the color bar in the upper-right side shows the $B-V$ color index ranges for the pseudo-colors.
\label{colmap}}
\end{figure}

All works in this paper have been carried out using the data set obtained by the Hubble Heritage Team. They observed M51 using the {\it HST} Advanced Camera for Surveys (ACS) with F435W, F555W, F814W and F658N filters as part of {\it HST} program 10452. The observation was completed in January 2005 and the data were publicly released in April 2005, covering about $6.8' \times 10.5'$ field centered on M51. The accumulated exposure times are 2720, 1360 and 1360 seconds in F435W, F555W and F814W, respectively. The basic data processing, multi-drizzling and image combination were done by the Space Telescope Science Institute (STScI) before the data release. More details on the observation and data reduction are available in \citet{mut05}.
In this paper, we used F435W, F555W and F814W (hereafter, we refer to these filters simply as $B$, $V$ and $I$, respectively)\footnote{Thus, in this paper, $BVI$ notations do not indicate the Johnson $BVI$ filters. } images.
Pixel surface brightness and its photometric error have been derived using the science and weight images\footnote{The M51 Hubble Heritage Program data and related information can be retrieved from the following webpage: http://archive.stsci.edu/prepds/m51/index.html}. Based on the instruction of the {\it HST} Data Handbook for ACS (version 5.0), we derived the following equations for pixel AB magnitude and photometric error in each band:
\begin{equation}
m_F = -2.5 \log P_F + Z_F,
\end{equation}
\begin{equation}
e_F = 2.5 \log e \times (P_F \times W_F)^{-0.5},
\end{equation}
where $m_F$, $e_F$, $P_F$ and $W_F$ are pixel AB magnitude, photometric error, science-image pixel value and weight-image pixel value in the $F$ band (with $F=B$, $V$ and $I$), respectively. $Z_F$ is magnitude zero point, with $Z_B=25.6732$, $Z_V=25.7349$ and $Z_I=25.9366$. Surface brightness is defined as AB magnitude per unit angular area (1 arcsec$^{-2}$).

The distance to NGC 5195 (i.e.~the distance to M51) was estimated by \citet{fel97} using the planetary nebula luminosity function, which we adopted in this paper: $8.4\pm0.6$ Mpc ($m-M=29.62\pm0.15$). This distance was confirmed by the recent study of \citet{vin12} using supernovae: $8.4\pm0.7$ Mpc. At this distance, the linear scale is 40 parsecs per arcsecond, which corresponds to 2.0 parsecs per pixel for the {\it HST}/ACS pixel size of $0.05''$. The foreground reddening toward M51 is $E(B-V)=0.035$, with the extinction in each band: $A_B=0.150$, $A_V=0.115$ and $A_I=0.067$ mag \citep{car89,sch98}. All magnitudes, surface brightnesses and colors in this paper have been corrected for the foreground extinction.

The angular pixel scale of the {\it HST}/ACS images is $0.05''$, but the typical full width at half maximum (FWHM) of a point source in the {\it HST}/ACS is about $0.1''$.
Thus, to ensure that each pixel is statistically independent of the surrounding pixels in the pixel analysis, it is necessary to bin the pixels by a factor of at least 2. In this paper, we binned the pixels by a factor of 4 ($4\times4$ binning) to improve the signal-to-noise ratio for each pixel.
After the binning, even very faint pixels have relatively good photometric quality (photometric error $\simlt0.1$ mag arcsec$^{-2}$).

Subsequent to the pixel analysis of NGC 5194 in Paper~I, we analyze NGC 5195 here. NGC 5195 seems to have the internal dust attenuation features different from that of NGC 5194, which makes the pixel properties of NGC 5195 very different. We divided NGC 5194 and NGC 5195 using a line that is vertical to the line connecting the centers of the two galaxies and by which the center-connecting line is divided with a distance ratio of 7:3, as shown in Figure~\ref{colmap}. The central coordinate (J2000) of NGC 5195 is RA = 13h 29m 59.54s and Dec = +47d 15m 58.3s, while that of NGC 5194 is RA = 13h 29m 52.72s and Dec = +47d 11m 43.4s, which are directly estimated using the {\it HST}/ACS $V$-band drizzled image.
The pixel analysis in this paper is carried out for the NGC 5195 area and the tidal bridge between the two galaxies.

\begin{figure*}[!t]
\plotone{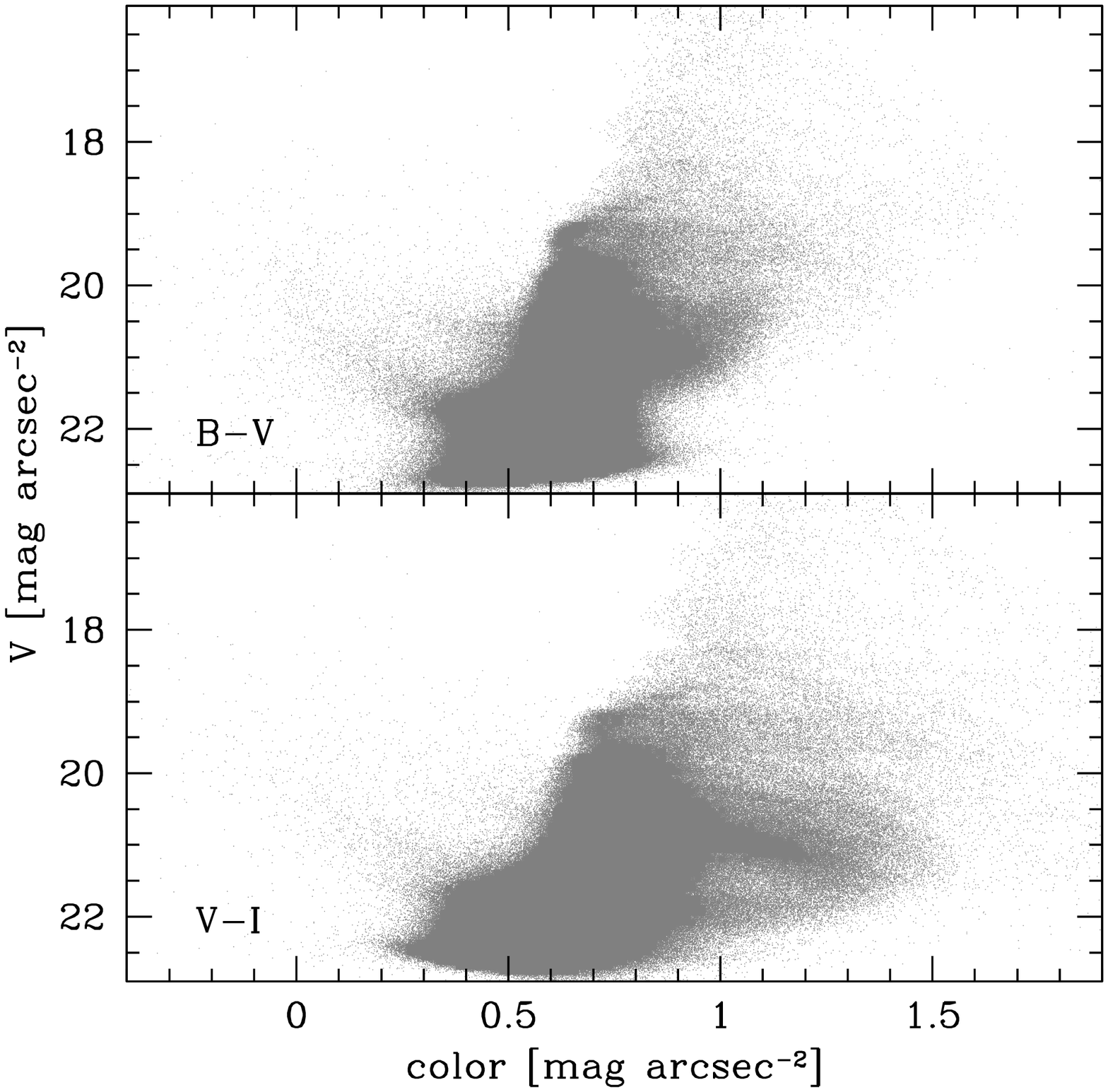}
\plotone{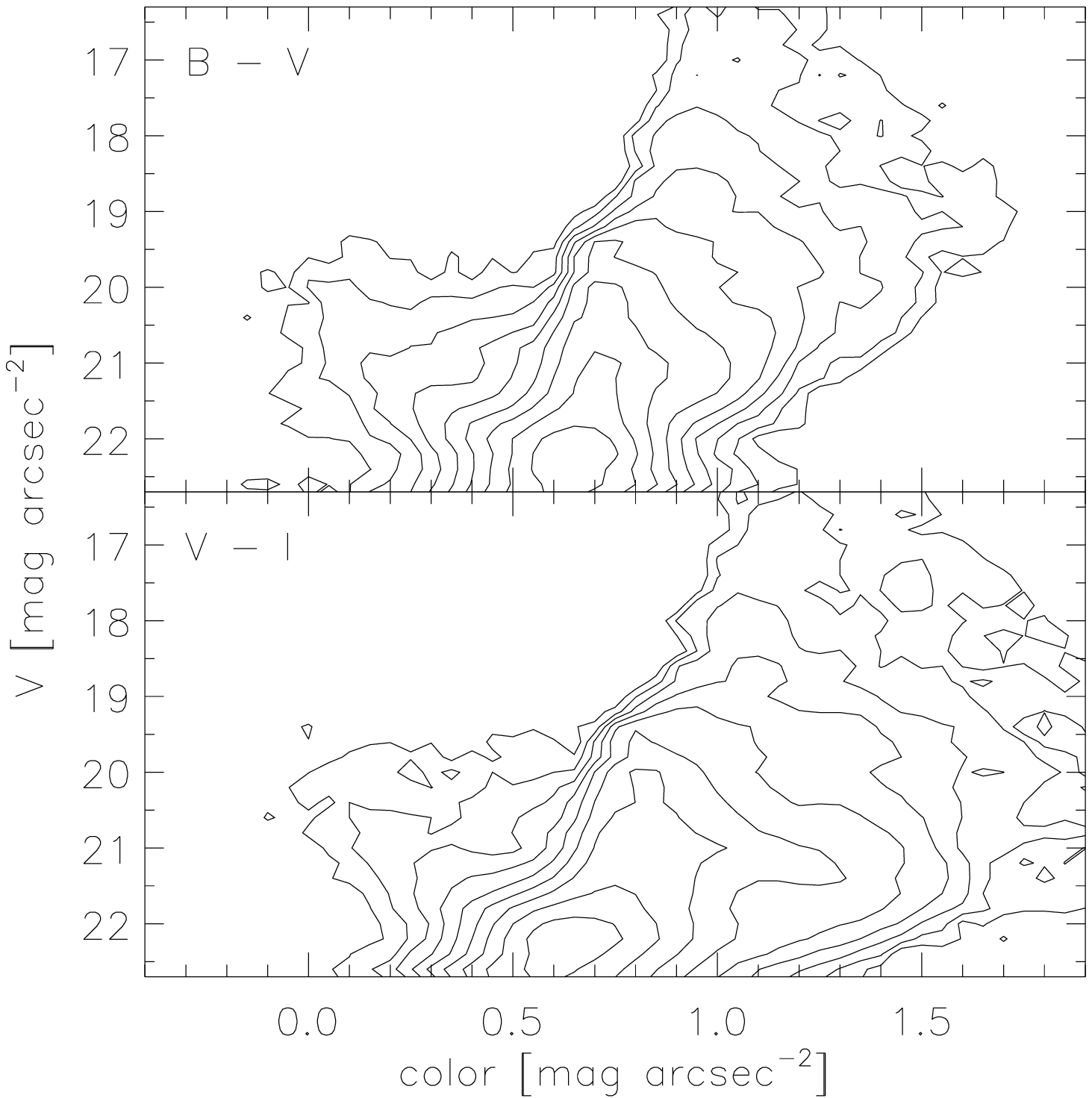}
\caption{ {\it Left panels}: pixel color versus magnitude diagrams (pCMDs) of NGC 5195 with $4\times4$ binning: $B-V$ versus $V$ (upper panel) and $V-I$ versus $V$ (lower panel). {\it Right panels}: contour maps showing the logarithmic pixel density distributions on the pCMDs of NGC 5195.
\label{allcmd}}
\end{figure*}

\section{RESULTS}\label{result}

\subsection{pCMD}

\subsubsection{Color distribution as a function of surface brightness}\label{cdist}

We first analyzed the pCMDs of NGC 5195 with $4\times4$ binning, which are shown in Figure~\ref{allcmd}. In those pCMDs, overall, brighter pixels tend to be redder, which is a trend similar to that of the red pixel sequence in NGC 5194 (Paper~I). However, the pCMDs of NGC5195 are clearly different from those of NGC 5194, in the sense that the blue pixel sequence is not as clear as that of NGC 5194. In the NGC 5195 pCMDs, most pixels seem to form a red pixel sequence with very large color dispersion and very small fraction of pixels form one or two small blue pixel branches at faint surface brightness (mostly $V\sim21$ mag arcsec$^{-2}$) \footnote{Here, `$V$' actually means the surface brightness in the $V$ band (i.e.~$\mu_V$). Throughout this paper, $\mu_V$ is simply denoted as $V$ for convenience.}.
The color dispersion of the `red pixel sequence' is very large and there are many very red pixels in the pCMDs. Such `very red pixels' are also found in NGC 5194 (Paper~I), but the amount of those pixels are much larger in NGC 5195. For example, at $20.5<V<21.0$ mag arcsec$^{-2}$, the reddest quarter of the NGC 5195 pixels have $V-I$ colors redder than 0.885, while those of the NGC 5194 pixels have $V-I$ colors redder than 0.317.

\begin{figure}[!t]
\plotone{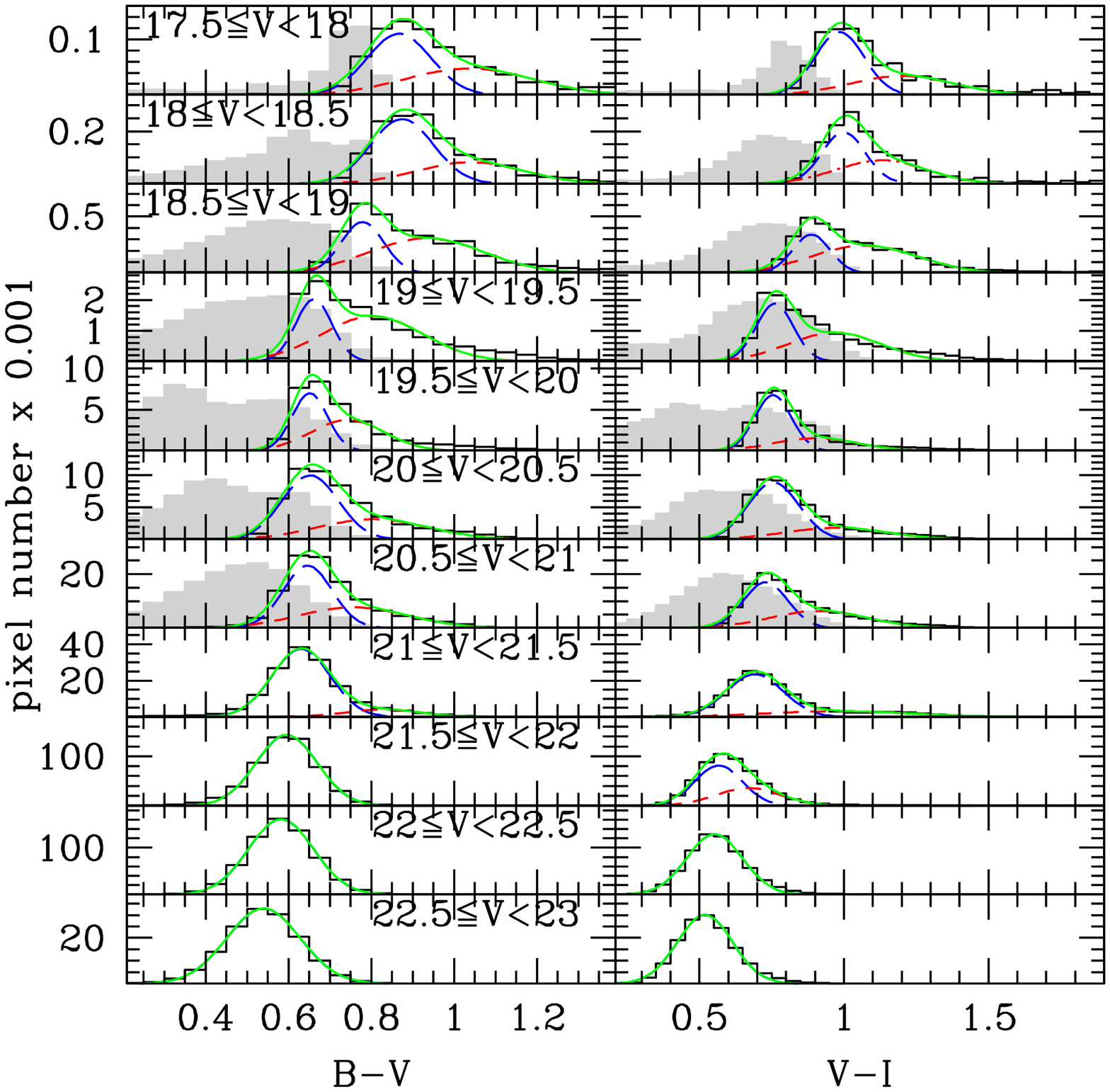}
\caption{ Color distributions as a function of pixel surface brightness with Gaussian fits. The dashed curves are individual Gaussians and the solid curves are the summed forms. For comparison, the pixel color distributions of NGC 5194 are overlaid as shaded histograms with adjusted scales.
\label{colfit}}
\end{figure}

For a quantitative analysis of the pCMD features, we investigated the color distribution as a function of pixel surface brightness. In Figure~\ref{colfit}, the pixel color distribution is presented for every 0.5 mag interval between $17.5\le V<23.0$ and fit using single or double Gaussian functions. The Gaussian fit parameters are summarized in Table~\ref{gauss}.
In Paper~I, we showed that the pixel color distribution of NGC 5194 is also fit well using double Gaussian functions at most surface brightness ranges (and rarely single or triple Gaussian functions). However, the physical implication of the double Gaussian components in NGC 5195 is quite different from that in NGC 5194. The double Gaussian components in the pCMD of NGC 5194 represent young disk pixels (blue pixel sequence) and old bulge pixels (red pixel sequence). On the other hand, the `blue' Gaussians in the NGC 5195 pCMDs do not represent disk pixels and actually they are not so blue. On average, the NGC 5195 `blue' Gaussians are redder than the NGC 5194 red Gaussians; for example, at $19.0\le V<19.5$ mag arcsec$^{-2}$, the NGC 5195 `blue' Gaussian peaks are $B-V=0.661$ and $V-I=0.761$, whereas the NGC 5194 red Gaussian peaks are $B-V=0.555$ $V-I=0.724$.
This indicates that most pixels in the NGC 5195 pCMD actually correspond to red sequence pixels. The `red' Gaussians in the NGC 5195 pCMD seem to be the result of a strong dust attenuation of the red sequence pixels. As described in the previous paragraph, the blue pixel sequence is so rarely populated in the NGC 5195 pCMD that those pixels are hardly found in the pixel color distribution plots.

\begin{figure}[!t]
\plotone{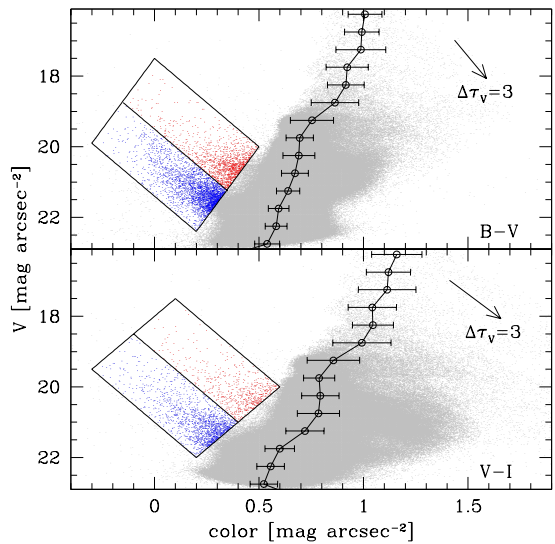}
\caption{ Pixel sequence definition on the $B-V$ (upper panel) and $V-I$ (lower panel) versus $V$ diagrams. Open circles show the median colors at the given $V$ surface brightness ranges and the errorbars indicate the sample inter-quartile ranges (SIQRs). The blue pixel sequences are manually defined, which are bordered with boxes. The arrows show the vectors of dust extinction with $\tau_V=3$.
\label{pcmd}}
\end{figure}

\begin{figure}[!t]
\plottwo{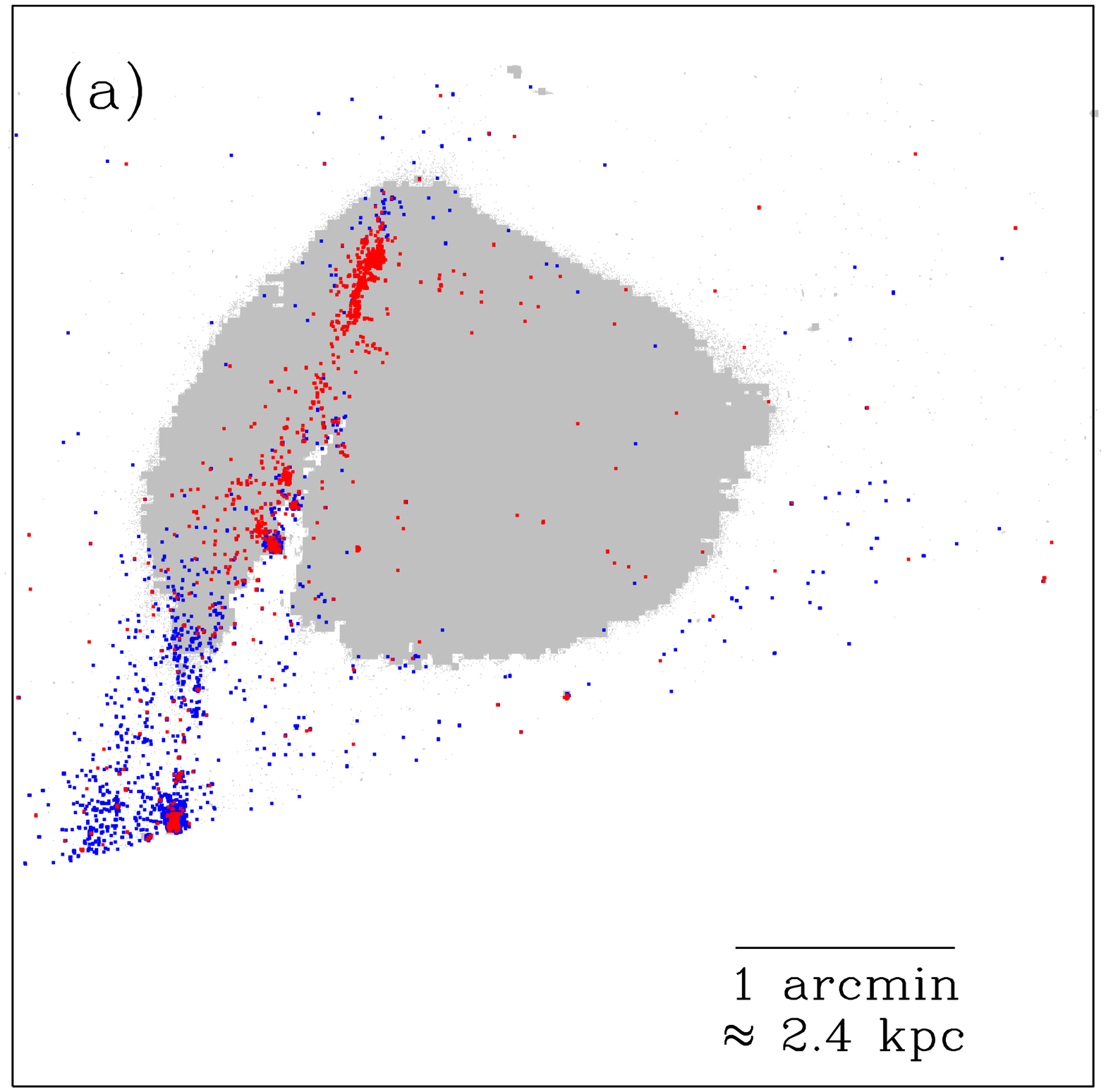}{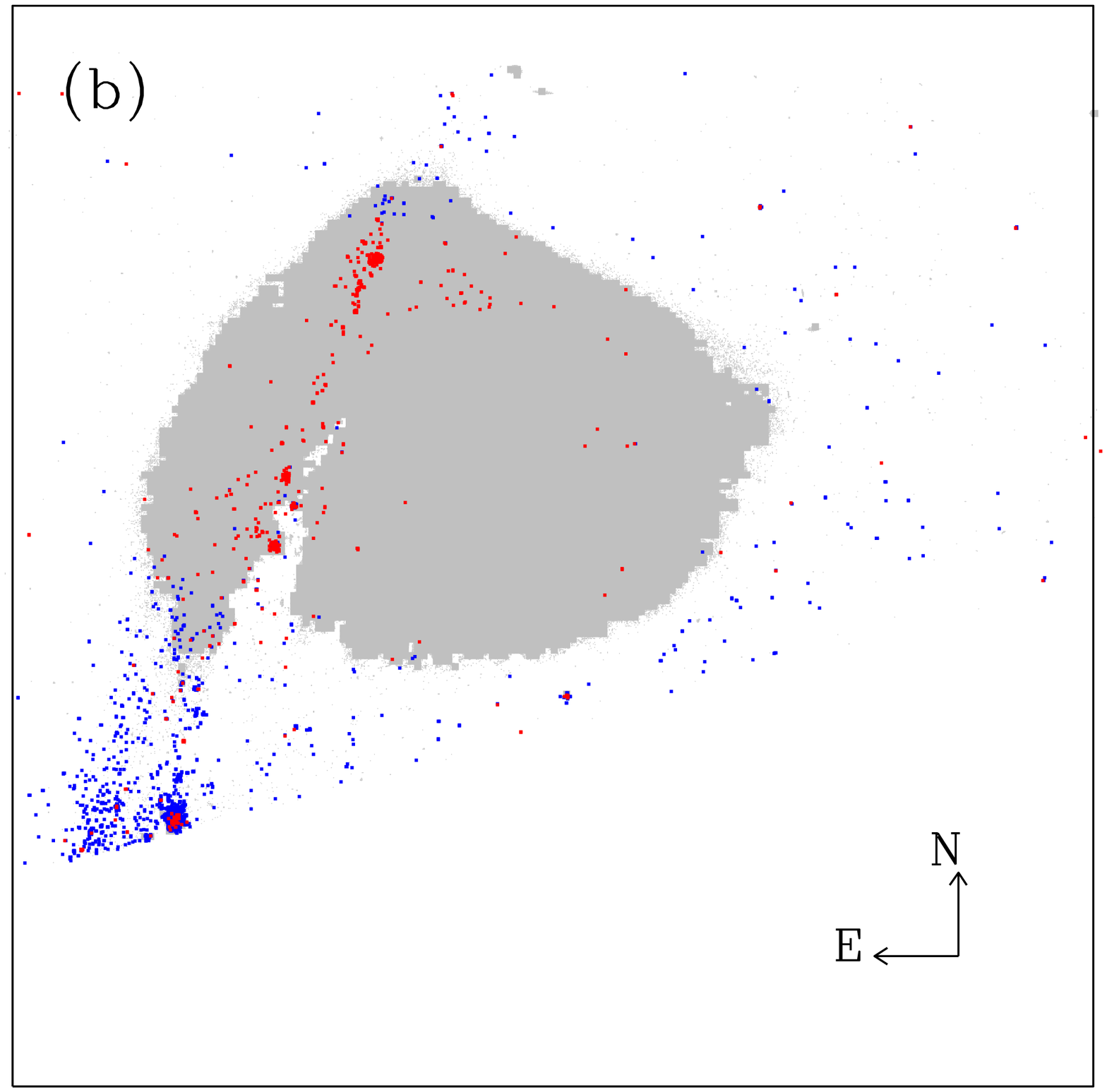}
\caption{ Spatial distributions of pixels ($V<21.5$ mag arcsec$^{-2}$) in the red pixel sequences (grey dots) and in the blue pixel sequence (blue and red dots; selected in Figure~\ref{pcmd}), defined using (a) $B-V$ and (b) $V-I$ pCMDs.
\label{cmap}}
\end{figure}

In Paper~I, we divided the NGC 5194 pixels into red and blue pixels, with fixed color cuts defined using the double Gaussian fits at $18.5\le V<19.5$ mag arcsec$^{-2}$, because the blue and red Gaussians show the best separation at that surface brightness range.
However, since there is no `actually blue' Gaussian in Figure~\ref{colfit}, we did not define fixed color cuts for NGC 5195. Instead, we manually selected the blue pixel sequences as shown in Figure~\ref{pcmd}. Except for those very small number of blue sequence pixels, most pixels of NGC 5195 seem to form a large red pixel sequence.
Figure~\ref{cmap} displays the spatial distribution of the pixels in the blue and red pixel sequences at $V<21.5$ mag arcsec$^{-2}$.
While the red sequence pixels are almost evenly distributed over NGC 5195, the blue sequence pixels are tightly gathered. The blue pixels form a stream-like feature, which is connected to the tidal bridge between NGC 5194 and NGC 5195, and weakly winds around NGC 5195.

It is noted that the spatial distributions of the two blue pixel sequences are obviously distinct. The pixels in the faint-blue-side blue pixel sequence are mostly clustered toward the middle point of the tidal bridge, while the pixels in the bright-red-side blue pixel sequence are relatively close to the NGC 5195 main body.
These spatial distributions show a possibility that the origin of the separated two blue pixel sequences is the difference in the dust extinction between the two locations (more details of the dust distribution in NGC 5195 is discussed in Section~\ref{svsp}).

In Figure~\ref{pcmd}, the median colors of the red sequence pixels as a function of surface brightness are overlaid. Since most pixels belong to the red pixel sequence, there is no actual difference between the median colors of the red sequence pixels and those of all the pixels. The red pixel sequence does not have a simple shape, but its slope obviously varies particularly in the $V$ versus $V-I$ pCMD. For example, the red sequence between $V\approx19.5$ and 21.0 is almost parallel to the Y-axis (magnitude-axis), while that between $V\approx18.0$ and 19.5 is significantly tilted.
These slope variations may be related to the transition between the structural components of NGC 5195 (e.g., bulge, bar and disk) with surface brightness, but any clear matching between surface brightness ranges and structural components is not found.
The red pixel sequence shows very large scatters in color, which may be due to the spread in the internal dust extinction. From the comparison with the dust extinction vector, the color dispersion of the red pixel sequence corresponds to $\Delta\tau_V\sim4-5$, where $\tau_V$ is the dust optical depth in the $V$ band.
The parameters describing the red pixel sequence are listed in Table~\ref{pcmdinfo}.

\subsubsection{Resolution dependence}

\begin{figure*}[!t]
\plotone{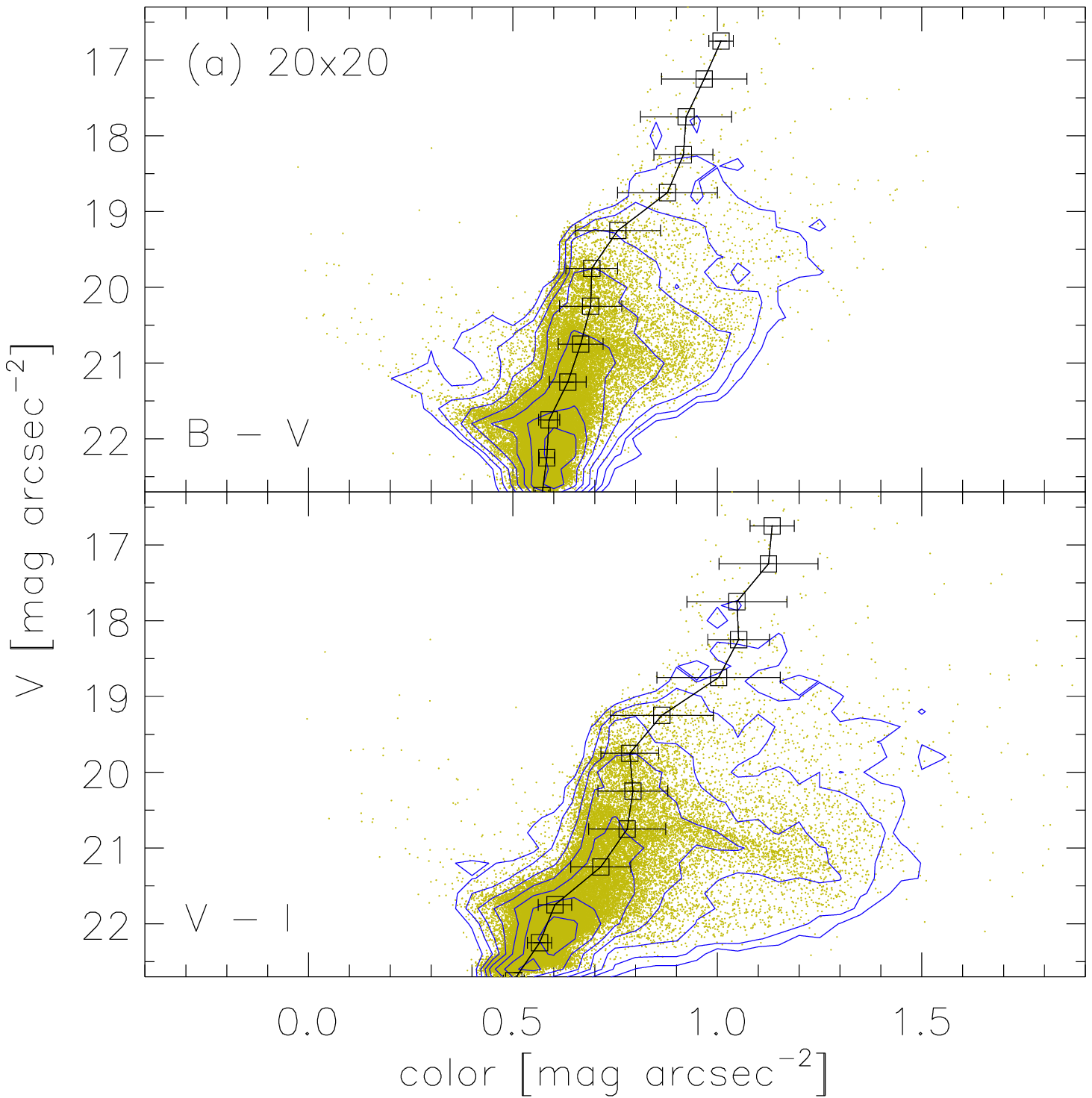}
\plotone{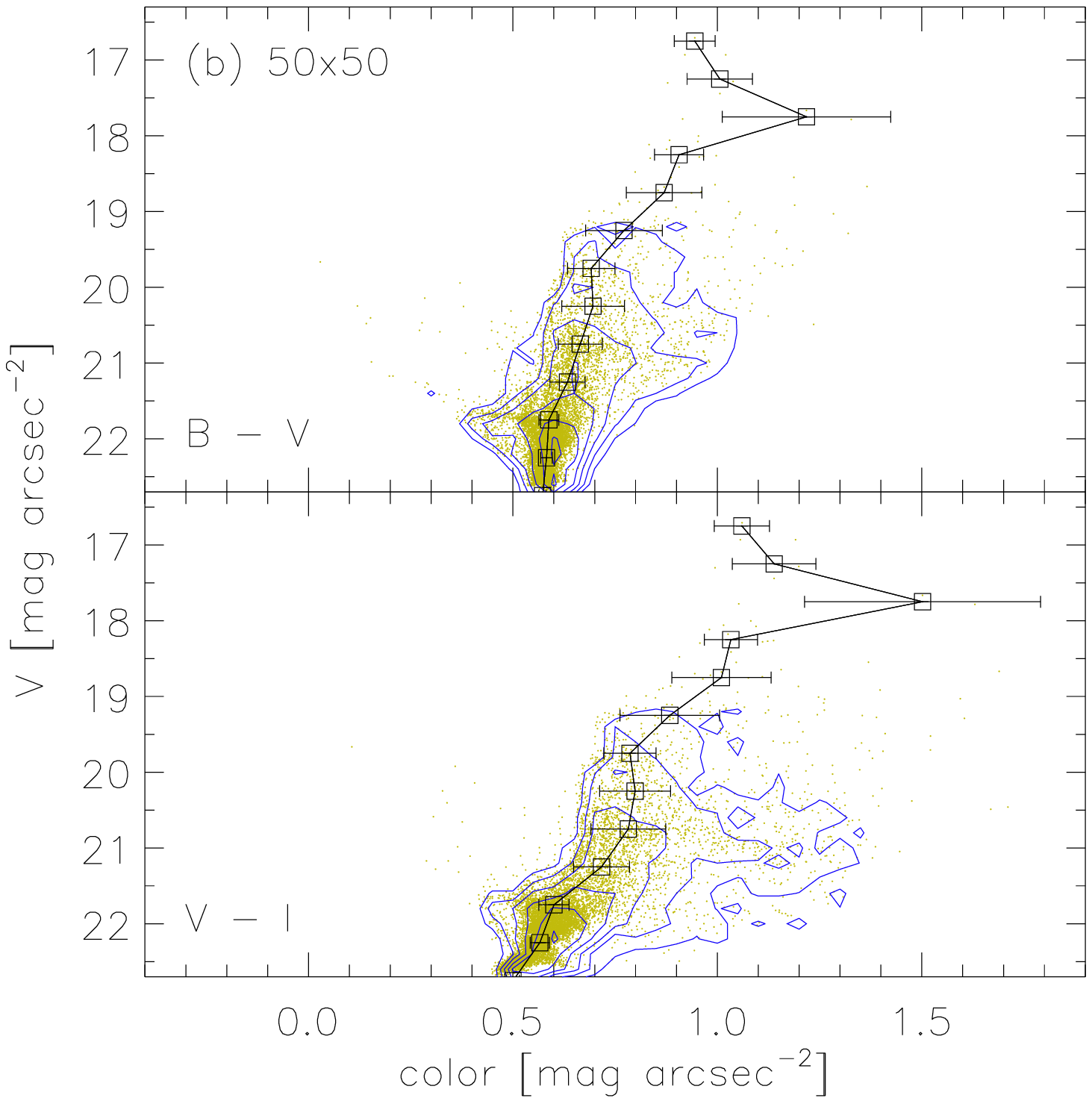}
\plotone{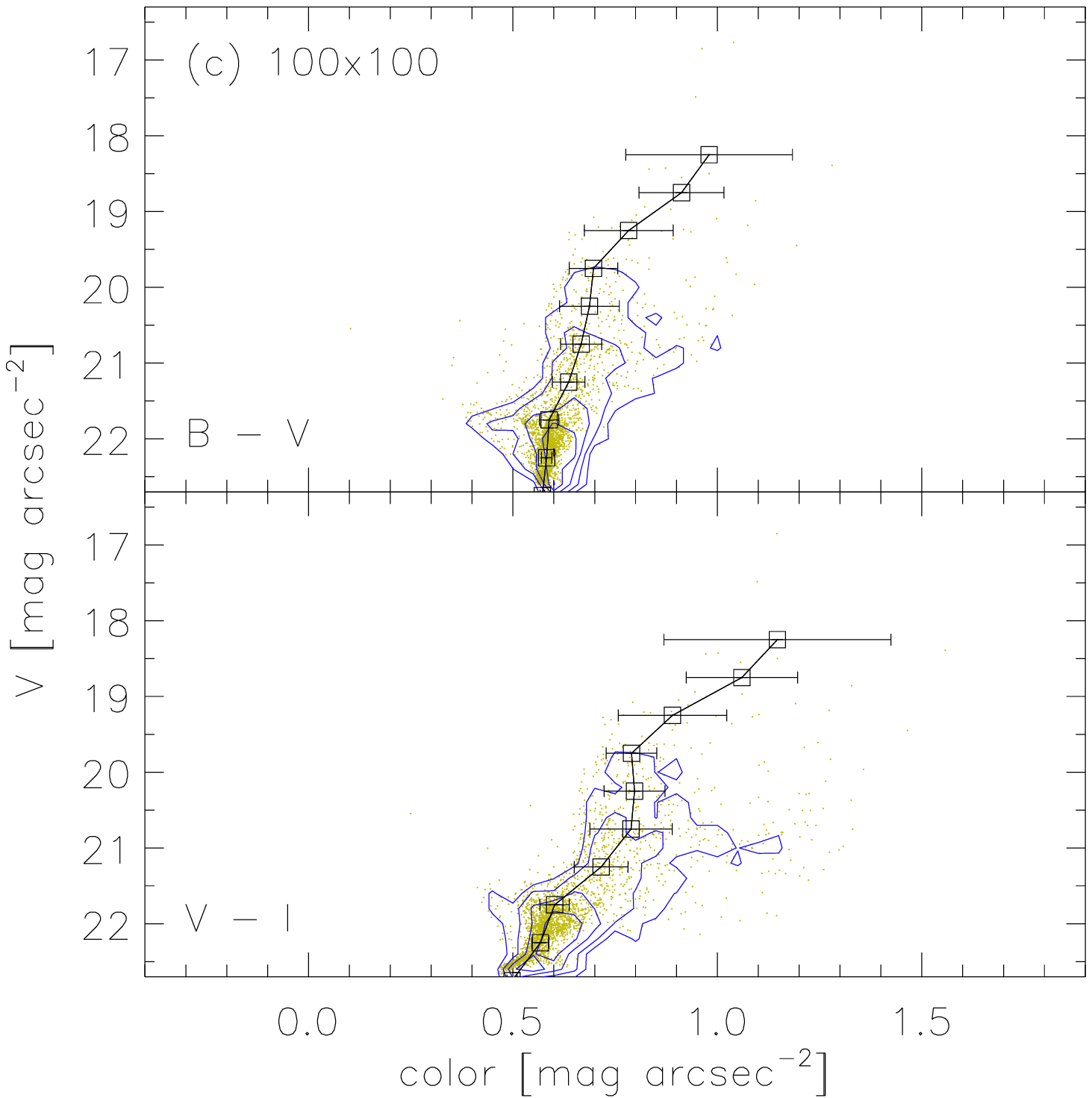}
\plotone{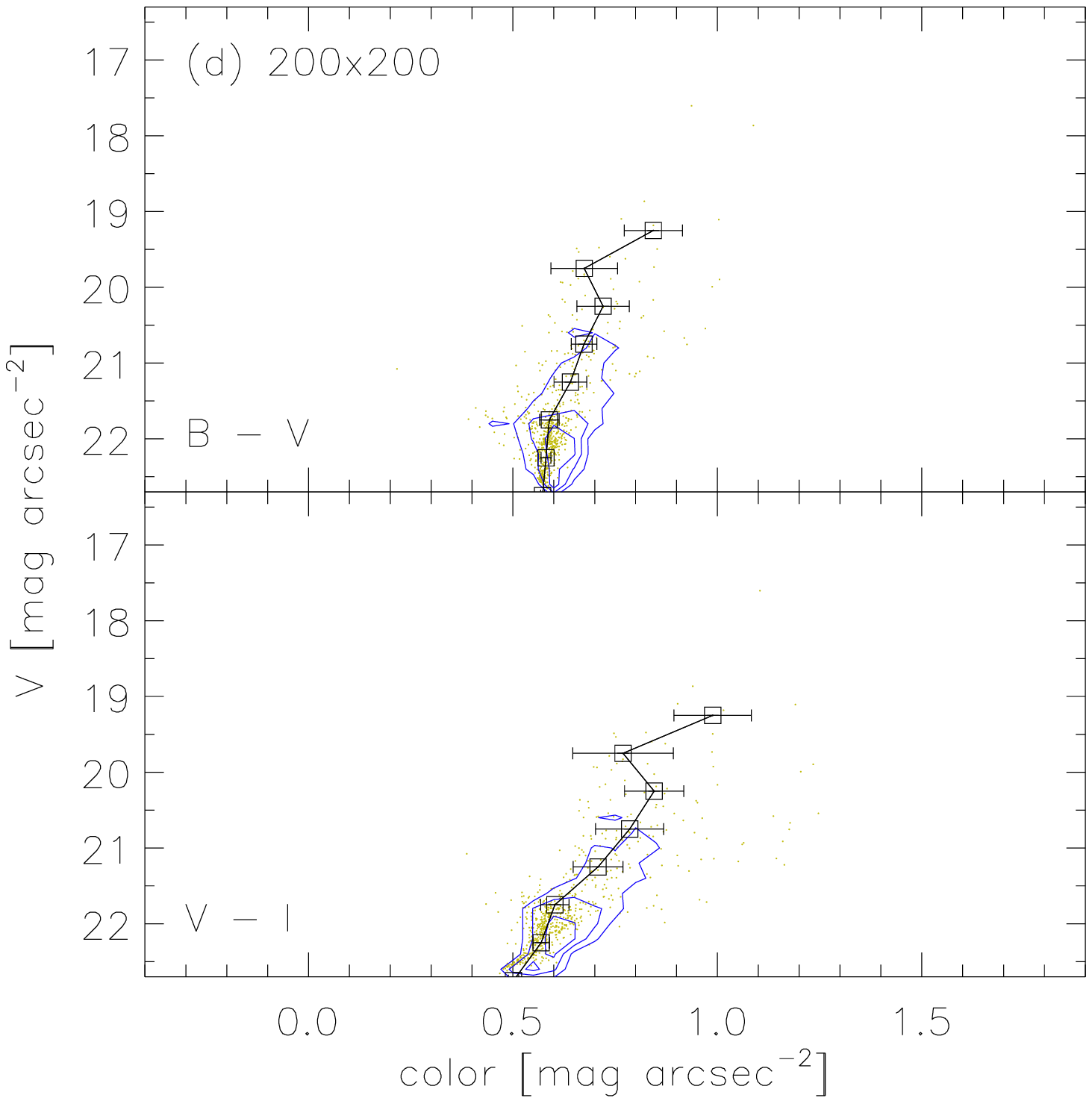}
\caption{ The pCMDs for different binning factors: (a) $20\times20$, (b) $50\times50$, (c) $100\times100$ and (d) $200\times200$ pixels.
Open rectangles show the median colors at the given $V$ surface brightness ranges and the errorbars indicate the SIQRs. The overlaid contour maps show the logarithmic pixel density distributions.
\label{pcmdbin}}
\end{figure*}

Since the pCMD features may vary as a function of image resolution, its resolution dependence needs to be checked.
For NGC 5194, in Paper~I, 100 pc pixel$^{-1}$ was a critical resolution, in the sense that important pCMD features are hardly distinguished in the resolutions worse than that. In Figure~\ref{pcmdbin}, we carried out a similar test for NGC 5195: the variation of the NGC 5195 pCMDs with different binning factors: $20\times20$, $50\times50$, $100\times100$ and $200\times200$ (corresponding to 40, 100, 200 and 400 pc pixel$^{-1}$ resolutions, respectively).

Like the trend for the NGC 5194 pCMDs, the color dispersion in the pCMD decreases as the spatial resolution becomes poor for the NGC 5195 pCMDs. The red pixel sequence does not seem to depend on spatial resolution significantly, while the blue pixel sequence is not distinguished at low resolution: recognizable marginally at the $100\times100$ binned pCMD and hardly at the $200\times200$ binned pCMD.
These results support the argument of Paper~I that the pCMD analysis requires at least 100 pc pixel$^{-1}$ resolution, which may correspond to the minimum scale from which multiple stellar components in galaxy-scale structures are significantly singled out.
Meanwhile, even in a resolution worse than 100 pc pixel$^{-1}$, the pCMD analysis of typical early-type galaxies (without interaction or current star formation) may be possible, because the entire shape of the red pixel sequence itself does not significantly affected by spatial resolution.
Table~\ref{pcmdinfo} lists the red pixel sequence parameters as a function of spatial resolution.

\subsection{Spatial Variation of Stellar Populations}\label{svsp}

\subsubsection{Stellar population division based on pCCD}

\begin{figure}[!t]
\plotone{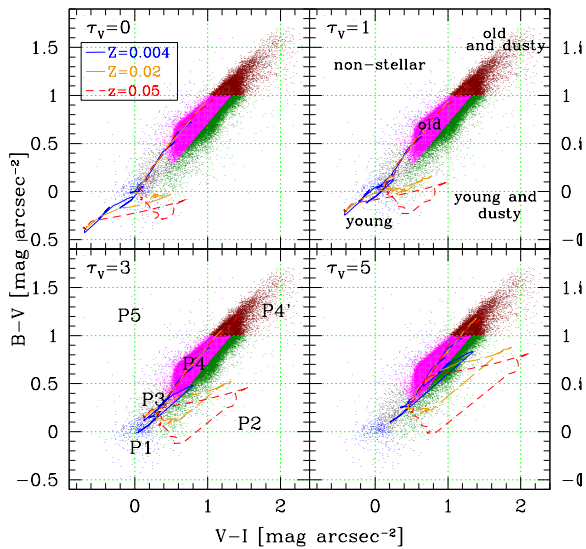}
\caption{ Pixel color-color diagrams (pCCDs) with simple stellar population (SSP) models \citep{bru03}. Dots are the pixels with $V<20$~mag~arcsec$^{-2}$ and the lines are SSP models with different metal abundances (Z = 0.02 is the solar abundance). Each panel shows the models with different total effective optical depth in the $V$ band ($\tau_V$) and fixed fraction of the ambient interstellar medium contribution ($\mu=0.3$), based on the simple two-component model of \citet{cha00}. The age of the SSP model at the lower-left end in each panel is about 0.1 Myr and that at the upper-right end is about 10 Gyr. The pixels are approximately divided into 6 populations: P1 (young), P2 (young and dusty), P3 (intermediate), P4 (old), P$4'$ (old and dusty) and P5 (non-stellar).
\label{pccd}}
\end{figure}

We investigated the spatial distribution of the NGC 5195 pixels, simply divided into several pixel populations.
In a $4\times4$ binned pixel ($\sim8\times8$ pc$^{2}$), the lights of many stars are integrated and the stars in each pixel may not be a simple stellar population (SSP) but a mixture of old stars and young stars. However, it is impossible to estimate the exact star formation history only using 3-band colors, and thus we suppose that the stars in a given pixel are an SSP to estimate their approximate age and metallicity by comparing their colors with population synthesis models. More details about this approximation are discussed in Paper~I.

Figure~\ref{pccd} shows the pCCD of NGC 5195 with population synthesis models \citep{bru03}. In Paper~I, we divided the NGC 5194 pixels into five pixel populations of P1 -- P5, which have different domains in the pCCD.
However, the pixel distribution in the pCCD of NGC 5195 seems to be significantly shifted from that of NGC 5194 (see also Figure 10 of Paper~I) toward increasing $B-V$ and $V-I$ colors. From the comparison with population synthesis models, this significant shift in the pCCD seems to have two causes: the lack of young blue stars and strong dust extinction for some old stars in NGC 5195.

For this reason, the five pixel populations defined in Paper~I are not enough to describe the pixel populations in NGC 5195. Thus, we added a new P4$'$ population, which is even redder than the red pixel population P4. Except for P4 and P$4'$, the definitions of the other pixel populations are the same as those in Paper~I. 
The population division criteria used in this paper are listed in Appendix \ref{crits}. 

A brief description with some approximation of physical quantities (age, metallicity and dust extinction) for each pixel population is as follows (see also Paper~I).
\begin{itemize}
\item[P1:] younger than 100 Myr and $\tau_V\sim0$.
\item[P2:] younger than 10 Myr and $\tau_V\gtrsim1$.
\item[P3:] approximately 100 Myr -- 1 Gyr old, but significantly degenerate in age, metallicity and dust attenuation.
\item[P4:] approximately 1 -- 10 Gyr old, but significantly degenerate in age, metallicity and dust attenuation.
\item[P$4'$]: older than 10 Gyr with $\tau_V\gtrsim3$.
\item[P5:] possibly contaminated by the light from extragalactic sources at high redshifts.
\end{itemize}
Without dust extinction, even very old ($\sim 10$ Gyr) populations can not have colors belonging to the P$4'$ domain.
For $\tau_V\sim5$, the P$4'$ pixels have SSP-equivalent ages $\gtrsim10$ Gyr with rich metallicity (Z $\gtrsim0.02$).
Due to the limit of the two-color-based classification, the pixel classification in Figure~\ref{pccd} can be used just as a rough approximation. Moreover, the significant dust extinction (which is inhomogeneously distributed) makes the physical interpretation of the pixel populations much more difficult. Nevertheless, the P1, P2 and P$4'$ pixels are thought to be relatively good approximations of young; young and dusty; and old and dusty stellar populations, respectively.
Thus, our analysis on the population variation mainly focuses on these three pixel populations.

\begin{figure}[!t]
\plotone{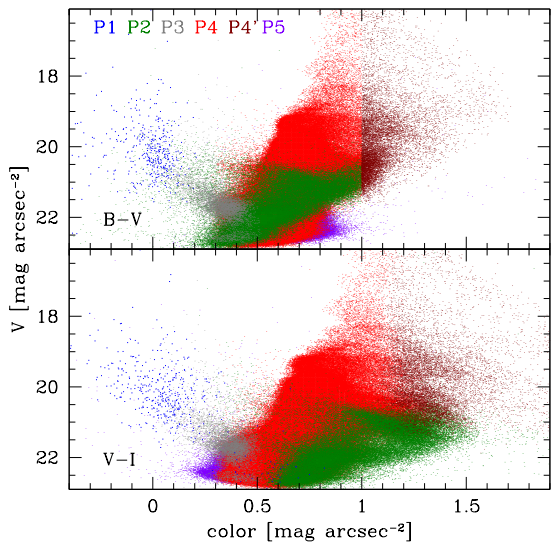}
\caption{ The pCMDs with the six populations defined in Figure~\ref{pccd}: P1 (blue), P2 (green), P3 (grey), P4 (red), P$4'$ (brown) and P5 (purple).
\label{pcmdpop}}
\end{figure}

Each population defined in the pCCD (Figure~\ref{pccd}) shows a characteristic distribution also in the pCMDs, as shown in Figure~\ref{pcmdpop}. The P1 (young) pixels form the weak blue pixel sequences in the pCMDs, while P4 (old) pixels form the main body of the red pixel sequence. The P3 (intermediate) pixels seem to connect the blue with red pixel sequences. It is noted that the two dusty populations (P2 and P$4'$) are defined just in the color-color domain (i.e.~not based on their brightness information), but their surface brightness domains are clearly distinct. That is, the surface brightnesses of the P2 pixels range over $20.0\lesssim V \lesssim23.0$ mag arcsec$^{-2}$, whereas those of P$4'$ pixels range over $16.0\lesssim V \lesssim21.5$ mag arcsec$^{-2}$.
The lack of young stars in NGC 5195 may be responsible for the low brightness of the P1 and P2 pixels, both of which are thought to mainly consist of young stars (either dusty or not).

\begin{figure}[!t]
\plotone{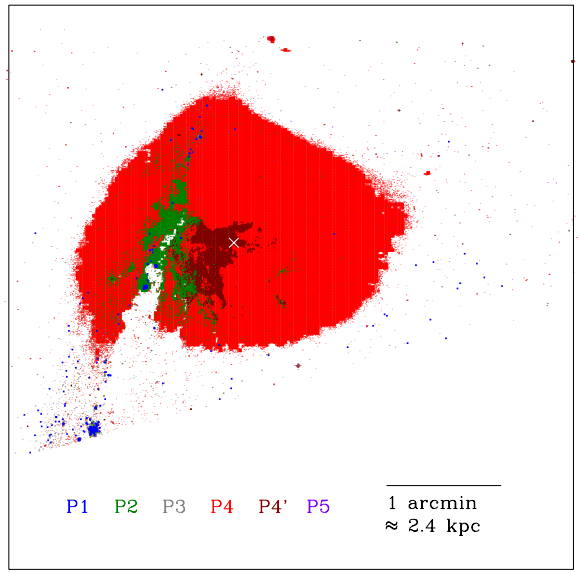}
\caption{ Spatial distribution of the six populations in NGC 5195, for pixels with $V<21.5$ mag arcsec$^{-2}$. The white cross is the NGC 5195 center: RA = 13h 29m 59.54s and Dec = +47d 15m 58.3s.
\label{popmap}}
\end{figure}

Figure~\ref{popmap} displays the spatial distributions of the six pixel populations classified in Figure~\ref{pccd}. The composition of pixel stellar populations strongly depend on the spatial location; particularly, the tidal bridge between NGC 5194 and NGC 5195 seems to affect significantly the pixel population composition.
The P1 pixels are mainly distributed along the tidal bridge and the outskirt of NGC 5195. The dusty pixels (P2 and P$4'$) also seem to be strongly affected by the tidal interaction between the two galaxies. The P2 pixels are distributed similarly to the P1 pixels (i.e.~along the tidal bridge), but gathered more closely to the NGC 5195 center. The P$4'$ pixels are spatially intimate to the P2 pixels and distributed between the tidal bridge and the NGC 5195 center. The overall distribution of the P2 and P$4'$ pixels indicates that there exists much dust in the area between the tidal bridge and the NGC 5195 center.

\subsubsection{NGC 5195 main body}\label{armdef}

\begin{figure}[!t]
\plotone{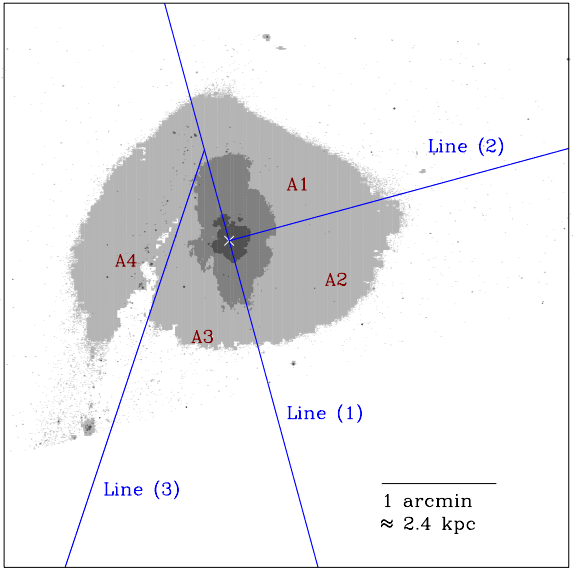}
\caption{ Divided areas for the comparison of populations. The line from upper-left to lower-right (Line (1))connects the centers of NGC 5194 and NGC 5195, to which the line between A1 and A2 (Line (2)) is perpendicular. The line between A3 and A4 (Line (3)) is manually selected to distinguish the tidal bridge area (A4). Dots with different darkness indicate the pixels with different surface brightness: light ($20.0\le V<21.5$ mag arcsec$^{-2}$), intermediate ($19.0\le V<20.0$ mag arcsec$^{-2}$) and dark ($V<19.0$ mag arcsec$^{-2}$). The white cross is the NGC 5195 center.
\label{areadiv}}
\end{figure}

\begin{figure}[!t]
\plotone{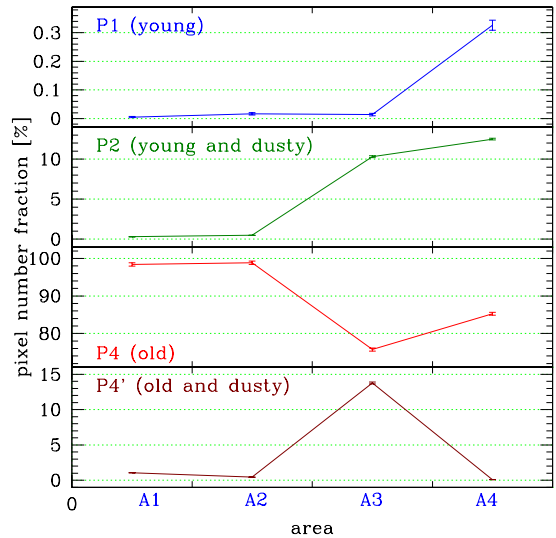}
\caption{ Population variations along the areas A1 -- A4, for P1 (young), P2 (young and dusty), P4 (old) and P$4'$ (old and dusty) populations with $V<21.5$ mag arcsec$^{-2}$. The errorbars indicate the Poisson errors.
\label{mbody}}
\end{figure}

To investigate the pixel populations as a function of location, we divided the NGC 5195 area into four sub-areas in Figure~\ref{areadiv}. For this division, we drew three lines: (1) the line connecting the centers of NGC 5194 and NGC 5195, (2) the line that starts from the NGC 5195 center and is perpendicular to Line (1), and (3) the line distinguishing the tidal bridge area (manually selected). The equations for these lines are as follows:
\begin{eqnarray}
\textrm{Line (1)}:&\delta-\delta_{0} = &3.679\; (\alpha - \alpha_{0}) ,\\
\textrm{Line (2)}:&\delta-\delta_{0} = &-0.272\; (\alpha - \alpha_{0}) ,\\
\textrm{Line (3)}:&\delta-\delta_{0} = &-3\; (\alpha - \alpha_{0}) + 87'',\label{bodyline}
\end{eqnarray}
where $\alpha$ and $\delta$ are RA and Dec in the unit of arcsec, and $\alpha_0$ and $\delta_0$ are the RA and Dec of the NGC 5195 center. The A3 area is toward the tidal bridge (A4), while the A1 and A2 areas are the opposite sides.

Figure~\ref{mbody} shows the population variation along the areas A1 -- A4. The P1 (young) fraction ($f_{P1}$) is less than $0.02\%$ in A1, A2 and A3, but it is as large as $0.3\%$ in A4 (larger by a factor of $>15$). The young and dusty population fraction ($f_{P2}$) is less than $0.5\%$ in A1 and A2, but it is larger than $10\%$ in A3 and A4 (larger by a factor of $>20$). The old population fraction ($f_{P4}$) is very large in A1 and A2 ($>98\%$), while it is relatively small in A3 ($\sim76\%$) and A4 ($\sim85\%$). The old and dusty population fraction  ($f_{P4'}$) is small ($\lesssim1\%$) in A1, A2 and A4, but it is as large as $14\%$ in A3 (larger by a factor of $>13$).
These results show that the fraction of young populations is significantly large in and around the tidal bridge and that the amount of dust is very large in the tidal-bridge-side main body of NGC 5195.

\subsubsection{Tidal bridge between NGC 5194 and NGC 5195}\label{popvartext}

\begin{figure}[!t]
\plotone{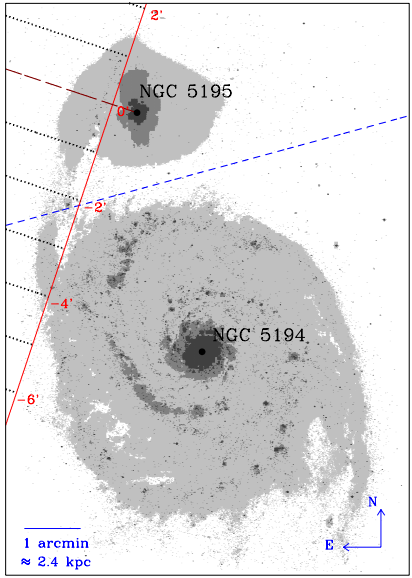}
\caption{ Tidal bridge area, which is defined as the upper-left area of Line (3). The solid line is Line (3) and the dashed line is the boundary between the NGC 5194 and NGC 5195 areas. The dotted lines are perpendicular to Line (3), denoting angular scales with one arcminute interval. The zero line (the long-dashed line) of the dotted lines is connected to the point closest to the NGC 5195 center, on Line (3).
\label{intarea}}
\end{figure}

\begin{figure}[!t]
\plotone{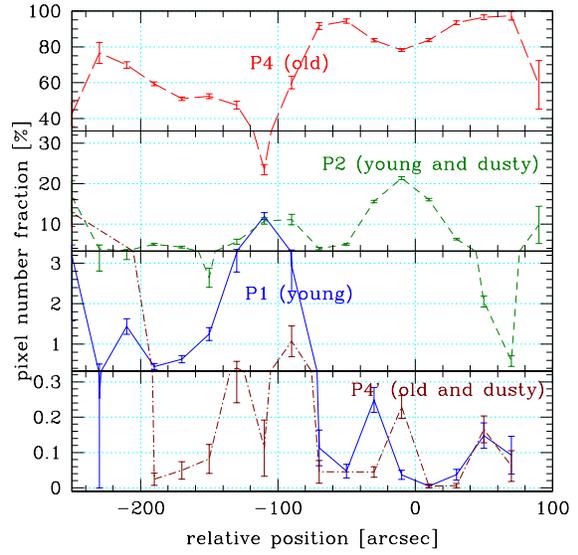}
\caption{ Population variations along the tidal bridge defined in Figure~\ref{intarea}, for P1 (solid), P2 (short-dashed), P4 (long-dashed) and P$4'$ (dot-dashed), with $V<21.5$ mag arcsec$^{-2}$. Note that the Y-axis is divided into four domains. The domain division is logarithmic, but each domain is in linear scale.
The errorbars indicate the Poisson errors.
\label{intpop}}
\end{figure}

In Figure~\ref{intarea} and Figure~\ref{intpop}, we investigated the pixel population variations along the tidal bridge between NGC 5194 and NGC 5195. Figure~\ref{intarea} shows the tidal bridge, which is defined as the north-east-side area of Line (3) (see Equation~\ref{bodyline}). The relative location along the tidal bridge is denoted as the angular separation ($\Delta\theta$) from the line penetrating the NGC 5195 center and perpendicular to Line (3).

With this area definition, Figure~\ref{intpop} shows the population variations along the tidal bridge. The most notable result in Figure~\ref{intpop} is the peak ($\sim12\%$) of the P1 population at $\Delta\theta\sim-110''$. This location corresponds to the middle point of the tidal bridge, where the tidal arms from NGC 5194 and NGC 5195 meet with each other, as shown in Figure~\ref{intarea}. It is interesting that the $f_{P1+P2}$ at $\Delta\theta\sim-10''$ is similar to that at $\Delta\theta\sim-110''$.
This means that young stars are clumped both at $\Delta\theta\sim-110''$ and $\Delta\theta\sim-10''$, but the tidal bridge middle point ($\Delta\theta\sim-110''$) has dust smaller than that around the NGC 5195 center ($\Delta\theta\sim-10''$).

\section{DISCUSSION}

\subsection{Interpretation of the pCMD}

\subsubsection{Blue pixel sequence}

Here, we discuss the physical implication of the pCMD features of NGC 5195 by comparing them with those of NGC 5194 shown in Paper~I.
The most conspicuous feature of the NGC 5195 pCMDs distinguished from the NGC 5194 pCMDs is that the NGC 5195 pCMDs have very weak blue sequences as shown in Figure~\ref{pcmd}. In addition, Figure~\ref{cmap} shows that the blue pixels (whose number is very small, compared to that of NGC 5194) form a stream-like feature in their spatial distribution.
At a first glance, the `blue pixel stream' seems to be the infall of young stars from NGC 5194 or the star formation regions along the infall of cold gas.

Recently, \citet{fon11} found a gas infall between two interacting galaxies NGC 5426 and NGC 5427, an event similar to which may have happened between NGC 5194 and NGC 5195. The gas flow from NGC 5194 to NGC 5195 is also found in hydrodynamic simulations \citep[e.g.,][]{dob10}. The blue pixels of NGC 5195 may trace this gas infall.
That is, the blue pixels of NGC 5195 are probably not an ordinary feature of a typical S0 galaxy, but a feature reflecting the tidal interaction between the two galaxies. Under the assumptions of Z = 0.02 and $\tau_V=1$, the estimated SSP-equivalent ages of the blue pixels range from 100 Myr to 1 Gyr, which indicates that the tidal interaction had been triggering star formation until very recent time.
However, there is no clear evidence that excludes the possibility that the young stars in the tidal bridge were moved from the spiral arm of NGC 5195 rather than were born in situ. One moderate speculation is that both of them may have happened: some of the young stars may have been moved from NGC 5195 and the other may have been formed from the gas inflow from NGC 5195.

\subsubsection{Red pixel sequence}

Rather than the blue pixel sequence, the red pixel sequence represents the ordinary feature of a typical S0 galaxy.
According to \citet{lan07}, the red pixel sequence is a principal feature to understand pCMDs of galaxies, particularly for early-type galaxies such as NGC 5195.
Most pixels of NGC 5195 belong to the red pixel sequence, whereas the blue pixels represent the transient properties triggered by the tidal interaction. Thus, the comparison of the red pixel sequences of NGC 5194 and NGC 5195 shows the difference between the two galaxies, in the relatively long-lasting properties.

The red pixel sequence in the NGC 5195 pCMD is systematically redder than that in the NGC 5194 pCMD. For example, the median colors of the NGC 5195 red pixel sequence at $V=19.5-20.0$ mag arcsec$^{-2}$ are $B-V=0.695$ and $V-I=0.788$, which are redder than those of the NGC 5194 red pixel sequence by $\Delta(B-V)=0.205$ and $\Delta(V-I)=0.115$. These differences vary with the surface brightness: $\Delta(B-V)=0.187$ and $\Delta(V-I)=0.248$ at $V=17.5-18.0$ mag arcsec$^{-2}$, and $\Delta(B-V)=0.171$ and $\Delta(V-I)=0.153$ at $V=20.5-21.0$ mag arcsec$^{-2}$, but it is clear that the red pixel sequence of NGC 5195 is significantly redder than that of NGC 5194 at any surface brightness.

\begin{figure}[!t]
\plotone{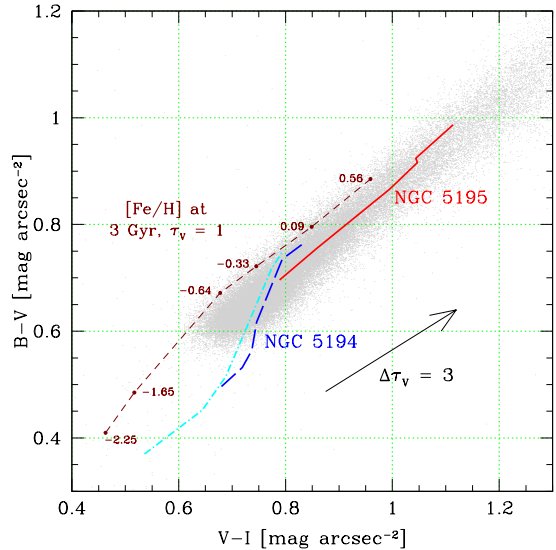}
\caption{ The red pixel sequence (= the total pixel sequence; $17<V<20$ mag arcsec$^{-2}$) of NGC 5195 (solid line) and the red and total pixel sequences of NGC 5194 (long-dashed and dot-dashed lines, respectively) on the $B-V$ versus $V-I$ diagram. The grey dots are the NGC 5195 pixels brighter than $V=20$ mag arcsec$^{-2}$.
The color variation along dust extinction is denoted as an arrow and that along metallicity variation is denoted as a short-dashed line. The [Fe/H] values under the assumptions of 3 Gyr age and $\tau_V=1$ are denoted along the short-dashed line.
\label{rsqccd}}
\end{figure}

In many previous studies, old stellar populations in the bulges of spiral galaxies are known to be as old as those in elliptical galaxies \citep[e.g.,][]{lee92,sar05,mac09}. Thus, it is a possible assumption that the difference in the mean age of red sequence pixels between NGC 5194 and NGC 5195 may not be so large to cause significant differences in their colors. With this assumption, the main origin of the color difference between the red pixel sequences of the two galaxies would be the difference in metallicity or dust.
Figure~\ref{rsqccd} compares the red pixel sequences of the two galaxies on the pCCD and shows how the colors depend on metallicity and dust. As shown in this figure, the directions of dust and metallicity variation vectors are very similar in the pCCD, which makes it difficult to distinguish the effects of metallicity and dust on the pixel colors.
If the metallicities of the stars in the red pixel sequences of NGC 5194 and NGC 5195 are not significantly different, the color difference in the red pixel sequence between NGC 5194 and NGC 5195 corresponds to the optical depth difference of $2<\Delta\tau_V<4$ by dust ($\sim2$ at $V=20$ mag arcsec$^{-2}$ and $\sim4$ at $V=17$ mag arcsec$^{-2}$).

From several recent infrared observations, the dust mass distribution in the M51 system has been investigated \citep[e.g.,][]{dra07}. In those results, the dust mass in NGC 5195 was estimated to be smaller than that in NGC 5194 by an order of 1 or 2. Since the stellar mass of NGC 5195 is known to be smaller than that of NGC 5194 at most by a factor of 2, the estimated dust mass shows that the total dust-to-stellar mass ratio of NGC 5195 is much smaller than that of NGC 5194. This may seem to be contradictory to the argument in the previous paragraph, but not necessarily, because most of the dust mass in NGC 5194 may exist in its disk and spiral arms rather than its bulge that is the main area for red pixels.
To confirm this problem, additional analysis of the detailed (pixel-scaled) distribution of dust mass using a high-resolution infrared data such as the {\it Herschel Space Observatory} \citep{pil10} data will be useful, although the spatial resolution in the far-infrared band is still much poorer than that in the optical band.

The red pixel sequence of NGC 5195 has a shape different from that of NGC 5194: the NGC 5195 red pixel sequence is almost linear in the pCCD, while the NGC 5194 red pixel sequence is a broken line. The shape of the total (red + blue) pixel sequence of NGC 5194 is also similar to that of the NGC 5194 red pixel sequence.
The color variations of the red pixel sequences in the NGC 5195 pCMDs (from $V=17$ mag arcsec$^{-2}$ to $V=20$ mag arcsec$^{-2}$) are $\frac{\Delta(B-V)}{\Delta V}\approx0.097$ and $\frac{\Delta(V-I)}{\Delta V}\approx0.107$, whereas those in the NGC 5194 pCMDs are $\frac{\Delta(B-V)}{\Delta V}\approx0.090$ and $\frac{\Delta(V-I)}{\Delta V}\approx0.053$. It is noted that the $\frac{\Delta(B-V)}{\Delta V}$ values are very similar between the two galaxies, but the $\frac{\Delta(V-I)}{\Delta V}$ value of NGC 5195 is twice larger than that of NGC 5194.

In Paper~I, we suggested two major factors determining the overall features of the red pixel sequence of a galaxy: metallicity gradient \citep[e.g.][]{ko05,lab09,tor10} and internal dust attenuation \citep[e.g.~red hook features;][]{lan07}.
The red pixel sequence is thought to be relatively insensitive to recent star formation, but it reflects metallicity and dust content. The red pixel sequence of NGC 5195 shows an excellent agreement with that argument in Figure~\ref{rsqccd}.
However, the red pixel sequence of NGC 5194 is not satisfactorily explained either by the metallicity effect or by the dust effect: neither of them agrees with the broken sequence of NGC 5194.
Currently, we have no clear idea about what causes the broken shape of the NGC 5194 red pixel sequence on the pCCD.
It may be due to the intrinsic difference between an early-type galaxy (NGC 5195) and the bulge of a spiral galaxy (NGC 5194), or it may be a unique property of NGC 5194 only. The pCCD comparisons of more late-type galaxies may be useful to conclude this.

\subsection{Influence of Tidal Interaction on Stellar Populations}

In Paper~I, it was shown that the pixel populations are not symmetrically distributed in the outer disk area ($R>2' \approx$ 4.8 kpc) of NGC 5194, in the sense that the young population fraction ($f_{P1}$) in the NGC 5195 side is larger than that in the opposite side by a factor of 2.3. This result indicates that the interaction enhances the star formation activity in the bridge area between NGC 5194 and NGC 5195.
This is not a surprising result, because the enhancement of star formation in the tidal bridge between two galaxies were already reported in some previous studies \citep[e.g.,][]{dem08,sen09}.

In this paper, we investigated the variation of pixel populations along the tidal bridge (Figure~\ref{intpop}), showing that the  $f_{P1+P2}$ shows a peak at the middle point of the tidal bridge, where the tidal arms from NGC 5194 and NGC 5195 seem to meet each other.
At this location, the $f_{P1}$ and $f_{P2}$ are almost the same as each other.
On the other hand, another peak of the $f_{P1+P2}$ is found near the center of NGC 5195, where most pixels with young stellar ages seem to be dusty (P2).
The implication of those population distributions around the tidal bridge is as follows. First, the young stars in the tidal bridge tend to be spatially clumped. This is in agreement with the diffuse ionized gas distribution traced using H{$\alpha$} and [S {\scriptsize II}] emission lines \citep{gre98} or using CO mapping \citep{kod11}, which shows bright gas clumps in the tidal bridge. The separation between the two peaks of $f_{P1+P2}$ is about 4 kpc and the young population fraction at the intermediate location between the two peaks is only $4\%$ (one fifth of the peak value). Second, the dust distribution does not agree with the young stars distribution. The dust fraction tends to increase with decreasing distance to the NGC 5195 center. Most young stars in the clump close to the NGC 5195 center seem to be dusty, whereas only a half of the young stars in the clump at $\Delta\theta\sim-110''$ from the NGC 5195 center do.

\section{CONCLUSION}

We carried out a pixel analysis of the interacting S0 galaxy NGC 5195, using the {\it HST}/ACS images in the $BVI$ bands.
After $4\times4$ binning of the {\it HST}/ACS images to secure a sufficient signal-to-noise ratio for each pixel, we have analyzed the pCMD and pCCD properties of NGC 5195. The spatial distributions of pixel stellar populations have been investigated, too.
We found several properties of NGC 5195, distinct from those of NGC 5194. Three main results were reported.

First, in the pCMDs of NGC 5195, most pixels form a large red pixel sequence with large scatters, which is thought to be the characteristics of an early-type (S0) galaxy. The very weak blue pixel sequences of NGC 5195 seem to be a transient feature, originating from the tidal interaction between NGC 5194 and NGC 5195. The estimated mean stellar ages of the blue pixels of NGC 5195 range from 100 Myr to 1 Gyr.

Second, the red pixel sequence of NGC 5195 is redder than that of NGC 5194, the difference of which corresponds to the dust optical depth difference of $2<\Delta\tau_V<4$ with an assumption of fixed age and metallicity. In addition, the scatter of the red pixel sequence is consistent with an internal dust extinction up to $\Delta\tau_V\approx5$. The dust distribution is a strong function of spatial location, in the sense that stronger dust extinction is found near the tidal bridge, which implies that this abundance of dust in NGC 5195 may be affected by the galaxy interaction. The shape of the red pixel sequence in the pCCD is different between NGC 5194 and NGC 5195. To check whether this is a typical difference between a late-type galaxy and an early-type galaxy or not, it is needed to compare pCMDs of more galaxies.

Finally, the area in and around the tidal bridge between NGC 5194 and NGC 5195 shows a significantly larger fraction of young stars than the other area of NGC 5195 (by a factor larger than 15), indicating that the galaxy interaction has triggered star formation or that many young stars have flowed in from the spiral arm of NGC 5194 via the interaction. The young stars (or young pixels) tend to be strongly clumped rather than evenly distributed along the tidal bridge, consistent with the clumpy distribution of ionized gas there. A young stars clump is found at the middle point of the tidal bridge.

On the other hand, the dusty population shows a relatively wide distribution around the tidal bridge and tends to be biased toward the NGC 5195 center. Since it is generally expected that cool gas and dust distributions are not very different \citep[e.g.,][]{ber83,kru94}, this disagreement between the spatial distributions of young stars and dust is an interesting issue, if those young stars have formed from the gas flowed in from NGC 5194. Currently, we do not have a clear explanation for this problem. An interesting topic for further studies may be to reproduce the spatial distributions of gas, dust and stars in M51 through numerical simulations.

In conclusion, the interaction with NGC 5194 seems to significantly affect the recent star formation history and dust enrichment of NGC 5195. Via the tidal bridge between the two galaxies, gas and dust may have fallen into NGC 5195, forming the blue stellar population and enriching the overall dust content of NGC 5195. This result shows that M51 is a good example of not only the galaxy-galaxy tidal interaction but also the hydrodynamic interaction between galaxies with different morphologies, which is thought to be an important driver of the morphological transformation of galaxies \citep{par09}.
As the pixel properties of an interacting early-type galaxy, the results in this paper will be compared with the pixel properties of typical non-interacting early-type galaxies in our future studies.

\acknowledgments

This paper presents the second results of the pilot studies for an observational research project, Pixel Analysis of Nearby Cluster Galaxies (PANCluG).
All authors are the members of Dedicated Researchers for Extragalactic AstronoMy (DREAM) in Korea Astronomy and Space Science Institute (KASI).

\begin{deluxetable}{crrrrrr}
\tablenum{1} \tablecolumns{7} \tablecaption{ Gaussian Fit Parameters of the Color Distributions as a Function of Pixel Surface Brightness\label{gauss}} \tablewidth{0pt}
\tablehead{$V$ range  & \multicolumn{3}{c}{Gaussian 1} & \multicolumn{3}{c}{Gaussian 2} \\
(mag arcsec$^{-2}$) & $A_1$ & $x_1$ & $\sigma_1$ & $A_2$ & $x_2$ & $\sigma_2$ }
\startdata
\multicolumn{7}{c}{(For $B-V$)} \\
17.5 -- 18.0 & 0.110 & 0.867 & 0.077 & 0.048 & 1.036 & 0.154 \\
18.0 -- 18.5 & 0.247 & 0.874 & 0.076 & 0.082 & 1.039 & 0.127 \\
18.5 -- 19.0 & 0.452 & 0.779 & 0.052 & 0.309 & 0.934 & 0.139 \\
19.0 -- 19.5 & 2.031 & 0.661 & 0.044 & 1.474 & 0.803 & 0.122 \\
19.5 -- 20.0 & 7.009 & 0.652 & 0.042 & 3.770 & 0.743 & 0.088 \\
20.0 -- 20.5 & 9.978 & 0.652 & 0.068 & 3.177 & 0.805 & 0.133 \\
20.5 -- 21.0 & 23.181 & 0.646 & 0.059 & 7.702 & 0.757 & 0.132 \\
21.0 -- 21.5 & 37.489 & 0.630 & 0.070 & 4.007 & 0.828 & 0.092 \\
21.5 -- 22.0 & 143.322 & 0.596 & 0.074 & --- & --- & --- \\
22.0 -- 22.5 & 160.725 & 0.582 & 0.079 & --- & --- & --- \\
22.5 -- 23.0 & 32.920 & 0.540 & 0.089 & --- & --- & --- \\
\hline
\\
\multicolumn{7}{c}{(For $V-I$)} \\
17.5 -- 18.0 & 0.113 & 0.985 & 0.085 & 0.035 & 1.203 & 0.178 \\
18.0 -- 18.5 & 0.196 & 0.999 & 0.075 & 0.090 & 1.131 & 0.155 \\
18.5 -- 19.0 & 0.342 & 0.887 & 0.065 & 0.244 & 1.071 & 0.186 \\
19.0 -- 19.5 & 1.883 & 0.761 & 0.065 & 0.938 & 0.968 & 0.160 \\
19.5 -- 20.0 & 6.801 & 0.755 & 0.063 & 1.554 & 0.904 & 0.145 \\
20.0 -- 20.5 & 8.912 & 0.756 & 0.088 & 1.835 & 0.973 & 0.178 \\
20.5 -- 21.0 & 17.015 & 0.730 & 0.076 & 6.386 & 0.913 & 0.167 \\
21.0 -- 21.5 & 23.421 & 0.692 & 0.098 & 3.122 & 0.979 & 0.254 \\
21.5 -- 22.0 & 81.209 & 0.566 & 0.082 & 35.540 & 0.670 & 0.112 \\
22.0 -- 22.5 & 128.673 & 0.553 & 0.096 & --- & --- & --- \\
22.5 -- 23.0 & 30.024 & 0.519 & 0.096 & --- & --- & --- \\
\enddata
\tablecomments{ The fitting function formula is as follows:\\
Fit($x$) $= A_1 \cdot \exp(-(x-x_1)^2/\,2\,{\sigma_1}^2) + A_2 \cdot \exp(-(x-x_2)^2/\,2\,{\sigma_2}^2) $, \\
where $x$ is the color index ($B-V$ or $V-I$).
}
\end{deluxetable}

\begin{deluxetable}{cccccc}
\tablenum{2} \tablecolumns{6} \tablecaption{ Red pixel sequences in the NGC 5195 pCMDs \label{pcmdinfo}} \tablewidth{0pt}
\tablehead{ Binning & $4\times4$ & $ 20\times20$ & $50\times50$ & $100\times100$ & $200\times200$ \\ $V$ range & Median (SIQR) & Median (SIQR) & Median (SIQR) & Median (SIQR) & Median (SIQR) }
\startdata
\multicolumn{6}{c}{(For $B-V$)} \\
16.0 -- 16.5 & $1.007$ (0.081) & 0.989 (0.091) & 1.067 (0.050) & --- & --- \\
16.5 -- 17.0 & $0.992$ (0.083) & 1.009 (0.030) & 0.945 (0.050) & --- & --- \\
17.0 -- 17.5 & $0.987$ (0.120) & 0.968 (0.104) & 1.006 (0.080) & --- & --- \\
17.5 -- 18.0 & $0.922$ (0.101) & 0.923 (0.111) & 1.218 (0.206) & --- & --- \\
18.0 -- 18.5 & $0.916$ (0.088) & 0.917 (0.072) & 0.906 (0.060) & 0.980 (0.204) & --- \\
18.5 -- 19.0 & $0.863$ (0.113) & 0.878 (0.122) & 0.870 (0.092) & 0.912 (0.104) & 0.821 (0.000) \\
19.0 -- 19.5 & $0.754$ (0.103) & 0.757 (0.104) & 0.772 (0.094) & 0.783 (0.109) & 0.844 (0.071) \\
19.5 -- 20.0 & $0.695$ (0.066) & 0.693 (0.063) & 0.691 (0.058) & 0.697 (0.059) & 0.674 (0.082) \\
20.0 -- 20.5 & $0.690$ (0.076) & 0.690 (0.076) & 0.696 (0.077) & 0.687 (0.073) & 0.721 (0.064) \\
20.5 -- 21.0 & $0.672$ (0.064) & 0.666 (0.055) & 0.665 (0.054) & 0.667 (0.050) & 0.674 (0.031) \\
21.0 -- 21.5 & $0.639$ (0.055) & 0.634 (0.045) & 0.633 (0.043) & 0.636 (0.040) & 0.641 (0.040) \\
21.5 -- 22.0 & $0.594$ (0.050) & 0.588 (0.026) & 0.588 (0.024) & 0.588 (0.024) & 0.589 (0.024) \\
22.0 -- 22.5 & $0.582$ (0.052) & 0.583 (0.019) & 0.582 (0.014) & 0.582 (0.013) & 0.581 (0.012) \\
22.5 -- 23.0 & $0.539$ (0.059) & 0.571 (0.016) & 0.572 (0.009) & 0.572 (0.005) & 0.572 (0.005) \\
\hline
\\
\multicolumn{6}{c}{(For $V-I$)} \\
16.0 -- 16.5 & $1.159$ (0.120) & 1.136 (0.136) & 1.186 (0.000) & --- & --- \\
16.5 -- 17.0 & $1.119$ (0.106) & 1.134 (0.054) & 1.060 (0.068) & --- & --- \\
17.0 -- 17.5 & $1.113$ (0.138) & 1.125 (0.121) & 1.139 (0.102) & --- & --- \\
17.5 -- 18.0 & $1.042$ (0.115) & 1.048 (0.122) & 1.502 (0.289) & --- & --- \\
18.0 -- 18.5 & $1.045$ (0.098) & 1.052 (0.076) & 1.033 (0.065) & 1.147 (0.278) & --- \\
18.5 -- 19.0 & $0.992$ (0.139) & 1.003 (0.151) & 1.010 (0.121) & 1.060 (0.136) & 0.940 (0.000) \\
19.0 -- 19.5 & $0.856$ (0.126) & 0.864 (0.126) & 0.884 (0.122) & 0.890 (0.133) & 0.989 (0.095) \\
19.5 -- 20.0 & $0.788$ (0.075) & 0.786 (0.071) & 0.786 (0.064) & 0.790 (0.062) & 0.769 (0.123) \\
20.0 -- 20.5 & $0.793$ (0.089) & 0.793 (0.085) & 0.799 (0.087) & 0.797 (0.074) & 0.845 (0.072) \\
20.5 -- 21.0 & $0.784$ (0.101) & 0.779 (0.094) & 0.782 (0.092) & 0.789 (0.101) & 0.785 (0.083) \\
21.0 -- 21.5 & $0.720$ (0.091) & 0.715 (0.073) & 0.717 (0.069) & 0.716 (0.066) & 0.708 (0.061) \\
21.5 -- 22.0 & $0.599$ (0.071) & 0.603 (0.041) & 0.600 (0.037) & 0.602 (0.036) & 0.603 (0.035) \\
22.0 -- 22.5 & $0.556$ (0.066) & 0.565 (0.030) & 0.566 (0.023) & 0.567 (0.021) & 0.569 (0.020) \\
22.5 -- 23.0 & $0.523$ (0.066) & 0.502 (0.026) & 0.500 (0.016) & 0.498 (0.012) & 0.501 (0.013) \\
\enddata
\end{deluxetable}

\begin{appendix}

\section{Criteria for the Pixel Population Division}\label{crits} 

We divided the pixel populations in this paper, using the following criteria:
\begin{eqnarray}
\lefteqn{\textrm{P1}: B-V < 0.8\times(V-I)+0.3}
\nonumber\\
& & \textrm{and } B-V < -(V-I)+0.2,\\
\lefteqn{\textrm{P2}: B-V < 0.9\times(V-I)-0.2}
\nonumber\\
& & \textrm{and } B-V \ge -(V-I)+0.2
\nonumber\\
& & \textrm{and } B-V < 1.0,\\
\lefteqn{\textrm{P3}: B-V \ge 0.9\times(V-I)-0.2}
\nonumber\\
& & \textrm{and } B-V < 0.8\times(V-I)+0.3
\nonumber\\
& & \textrm{and } B-V \ge -(V-I)+0.2
\nonumber\\
& & \textrm{and } B-V < -(V-I)+0.8,\\
\lefteqn{\textrm{P4}: B-V \ge 0.9\times(V-I)-0.2}
\nonumber\\
& & \textrm{and } B-V < 0.8\times(V-I)+0.3
\nonumber\\
& & \textrm{and } B-V \ge -(V-I)+0.8
\nonumber\\
& & \textrm{and } B-V < 1.0,\\
\lefteqn{\textrm{P}4': B-V < 0.8\times(V-I)+0.3}
\nonumber\\
& & \textrm{and } B-V \ge 1.0,\\
\lefteqn{\textrm{P5}: B-V \ge 0.8\times(V-I)+0.3.}
\end{eqnarray}

\end{appendix}

\end{document}